\documentclass[12pt]{article}
\usepackage[dvips]{graphicx}
\usepackage{epsfig}
\usepackage{amsmath,amssymb,amsbsy}
\usepackage{amsfonts}

\begin{document}
\thispagestyle{empty}

\begin{center}
{\Large\bf{
NEW RESULTS IN THE QUANTUM \\STATISTICAL APPROACH TO PARTON
\vskip 0.3cm
DISTRIBUTIONS} \footnote {Invited talk presented by J. Soffer at the "`QCD Evolution Workshop"', May 12 - 16, 2014, Santa Fe, New Mexico, USA (to be published in World Scientific Conference Proceedings)}}
\vskip 0.8cm
{\bf JACQUES SOFFER}

Physics Department, Temple University,\\
1835 N, 12th Street, Philadelphia, PA 19122-6082, USA\\
E-mail: jacques.soffer@gmail.com
\vskip 0.3cm
{\bf CLAUDE BOURRELY}

Aix-Marseille Universit\'e, D\'epartement de Physique,\\
Facult\'e des Sciences site de Luminy, 13288 Marseille, Cedex 09, France\\
E-mail: bourrely@cmi.univ-mrs.fr
\vskip 0.3cm

{\bf FRANCO BUCCELLA}

INFN, Sezione di Napoli,\\
Via Cintia, Napoli, I-80126, Italy\\
E-mail: buccella@na.infn.it

\vskip 0.8cm
{\bf Abstract}\end{center}
We will describe the quantum statistical approach to parton distributions
allowing to obtain simultaneously the unpolarized distributions and the
helicity distributions. We will present some recent results, in particular
related to the nucleon spin structure in QCD. Future measurements are
challenging to check the validity of this novel physical framework.


\newpage

\section{Basic review on the statistical description}
Let us first recall some of the basic components for building up the parton
distribution functions (PDF) in the statistical approach, as oppose to the
standard polynomial type parametrizations, based on Regge theory at low $x$ and
counting rules at large $x$. The fermion distributions are given by the sum of
two terms \cite{bbs1}, the
first one,
a quasi Fermi-Dirac function and the second one, a flavor and helicity
independent diffractive
contribution equal for light quarks. So we have, at the input energy scale
$Q_0^2$,
\begin{equation}
xq^h(x,Q^2_0)=
\frac{AX^h_{0q}x^b}{\exp [(x-X^h_{0q})/\bar{x}]+1}+
\frac{\tilde{A}x^{\tilde{b}}}{\exp(x/\bar{x})+1}~,
\label{eq1}
\end{equation}
\begin{equation}
x\bar{q}^h(x,Q^2_0)=
\frac{{\bar A}(X^{-h}_{0q})^{-1}x^{\bar b}}{\exp [(x+X^{-h}_{0q})/\bar{x}]+1}+
\frac{\tilde{A}x^{\tilde{b}}}{\exp(x/\bar{x})+1}~.
\label{eq2}
\end{equation}
It is important to remark that $x$ is indeed the natural variable, and not the
energy like in statistical mechanics, since all sum rules we will use are expressed
in terms of $x$.
Notice the change of sign of the potentials
and helicity for the antiquarks.
The parameter $\bar{x}$ plays the role of a {\it universal temperature}
and $X^{\pm}_{0q}$ are the two {\it thermodynamical potentials} of the quark
$q$, with helicity $h=\pm$. We would like to stress that the diffractive
contribution
occurs only in the unpolarized distributions $q(x)= q_{+}(x)+q_{-}(x)$ and it
is absent in the valence $q_{v}(x)= q(x) - \bar {q}(x)$ and in the helicity
distributions $\Delta q(x) = q_{+}(x)-q_{-}(x)$ (similarly for antiquarks).
The {\it nine} free parameters \footnote{$A$ and $\bar{A}$ are fixed by the
following normalization conditions $u-\bar{u}=2$, $d-\bar{d}=1$.} to describe
the light quark sector ($u$ and $d$), namely $X_{u}^{\pm}$, $X_{d}^{\pm}$, $b$,
$\bar b$, $\tilde b$, $\tilde A$ and $\bar x$
in the above expressions, were
determined at the input scale from the comparison with a selected set of
very precise unpolarized and polarized Deep Inelastic Scattering (DIS) data
\cite{bbs1}. The additional factors $X_{q}^{\pm}$ and $(X_{q}^{\pm})^{-1}$ come
from the transverse momentum dependence (TMD), as explained in Refs.~[2,3] (See
below). For the gluons we consider the black-body inspired expression
\begin{equation}
xG(x,Q^2_0)= \frac{A_Gx^{b_G}}{\exp(x/\bar{x})-1}~,
\label{eq5}
\end{equation}
a quasi Bose-Einstein function, with $b_G$, the only free parameter, since
$A_G$ is determined
by the momentum sum rule. We also assume a similar expression for the polarized
gluon
distribution
$x\Delta G(x,Q^2_0)={\tilde A}_Gx^{{\tilde b}_G}/[\exp(x/\bar{x})-1]$.
For the strange quark
distributions, the simple choice made in Ref.~[1]
was greatly improved in Ref.~[4]. Our procedure allows to construct
simultaneously the unpolarized quark distributions and the helicity
distributions. This is worth noting because it is a very unique situation.
Following our first
paper in 2002, new tests against experimental (unpolarized and
polarized) data turned out to be very satisfactory, in particular in hadronic
collisions, as reported in Refs.~[5,6].

\section{Some selected recent preliminary results}
Since 2002 a lot of new DIS data have been published and although our early
determination of the PDF has been rather successful, which reflects the fact
this physical approach lies on solid grounds, we felt that it was timely to
revisit it.\\
We have slightly increased the number of free parameters, in particular to
describe the strange quark distributions, and these parameters were determined
from a next-to leading order (NLO) fit of
a large set of accurate DIS data, (the unpolarized structure functions
$F_2^{p,n,d}(x,Q^2)$, the polarized structure functions $g_1^{p,n,d}(x,Q^2)$,
the structure function $xF_3^{\nu N}(x,Q^2)$ in $\nu N$ DIS, etc...) a total of
2140 experimental points. Although the full details of these new results in
their final form will be presented in a forthcoming paper \cite{bbs7}, we just
want to make a general remark. By comparing with the results of 2002
\cite{bbs1}, we have observed, so far, a remarquable stability of some
important parameters, the light quarks potentials $X_{0u}^{\pm}$ and
$X_{0d}^{\pm}$, whose numerical values are almost unchanged. The new
temperature is slightly lower. As a result the main features of the new light
quark and antiquark distributions are only hardly modified, which is not
surprizing, since our 2002 PDF set has proven to have a rather good predictive
power.

\begin{figure}[ht]
\centering
\includegraphics[width=6.2cm]{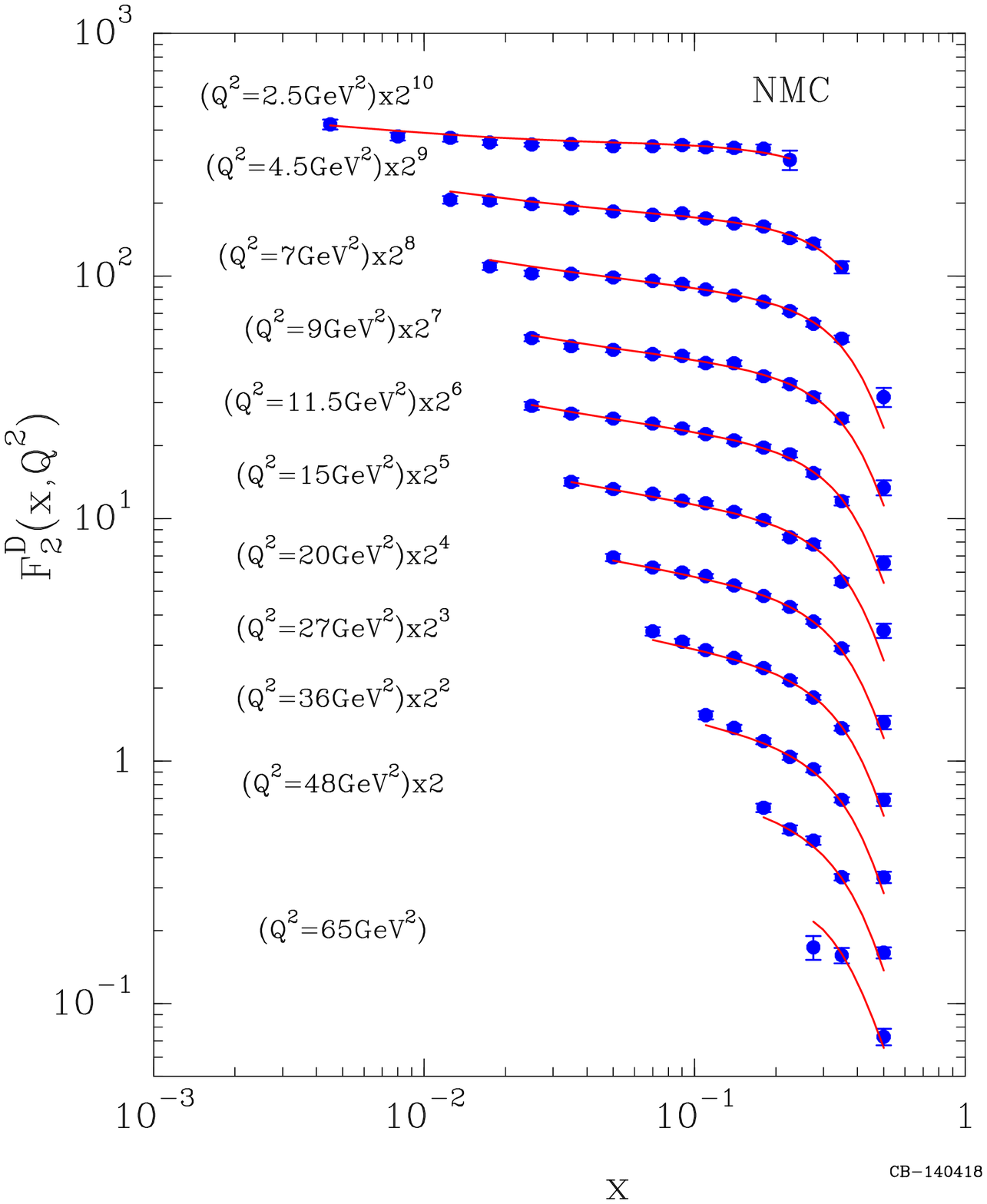}
\includegraphics[width=6.2cm]{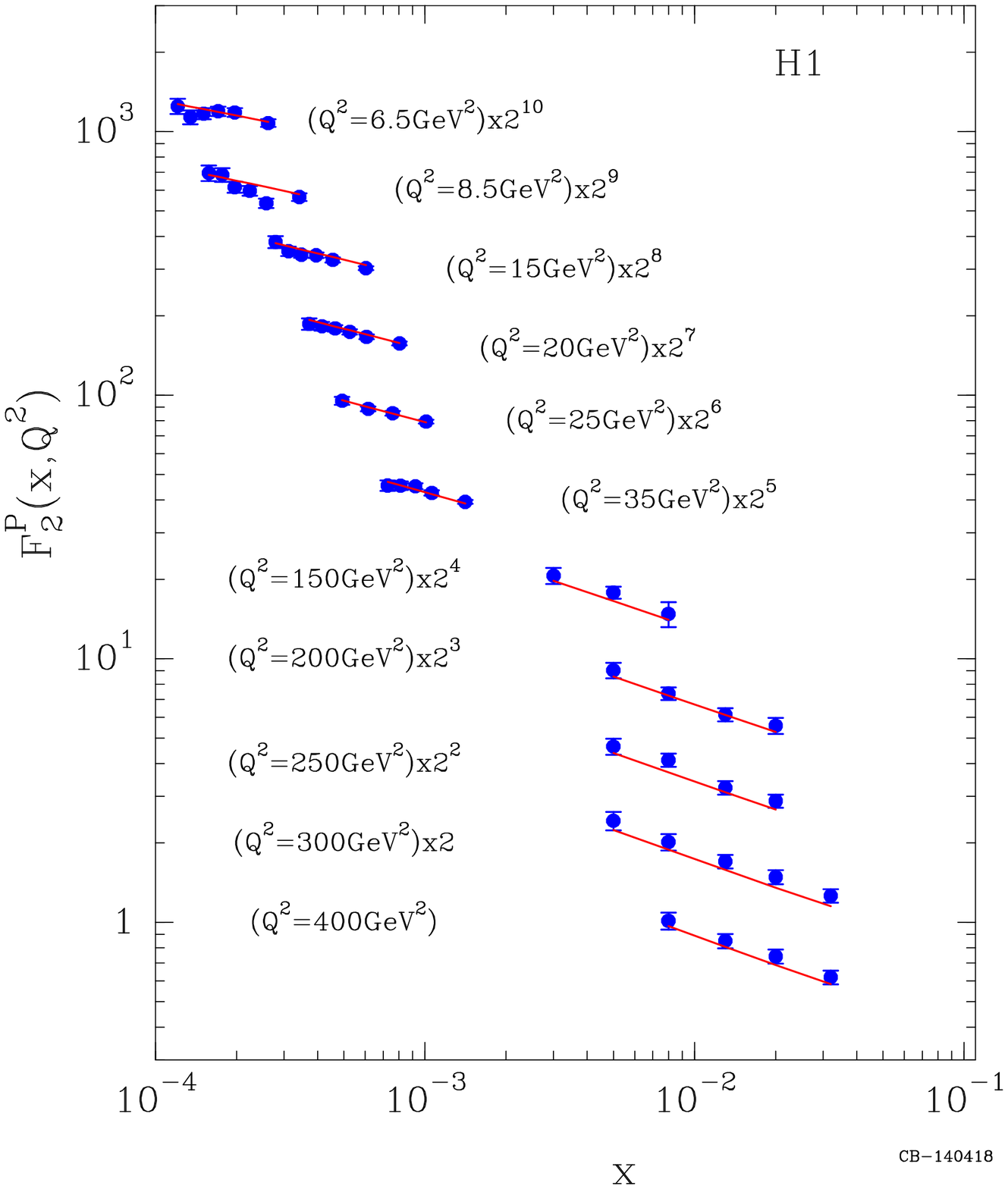}
\caption{ {\it Left}~: Comparison of the data on $F_2 ^D (x,Q^2)$  from NMC
[8],
with the predictions of the statistical model (solid curves).
{\it Right}~: Comparison of some selected data on $F_2^p (x,Q^2)$  from H1 [9],
with the predictions of the statistical model (solid curves).}
\label{fig1}
\end{figure}
First we present some selected experimental tests for the unpolarized PDF by
considering $\mu N$ and $e N$ DIS, for which several experiments have yielded a
large number of data points on the structure functions $F_2^N(x,Q^2)$, $N$
stands for either a proton or a deuterium target. We have used fixed target
measurements which cover a rather limited kinematic region in $Q^2$ and $x$ and
also HERA data which cover a very large $Q^2$ range and probe the very low $x$
region, dominated by a fast rising  behavior, consistent with our diffractive
term (See Eq. (1)).\\
For illustration of the quality of our fit and, as an example, we show in
Fig.~\ref{fig1}, our results for $F_2 ^D (x,Q^2)$ with NMC data on a deuterium
target and for $F_2 ^p (x,Q^2)$ with H1 data on a proton target. We note that
the analysis of the scaling violations leads to a gluon distribution
$xG(x,Q^2)$, in fairly good agreement with our simple parametrization (See Eq.
(3)).

\begin{figure}[ht]
\centering
\includegraphics[width=6.2cm]{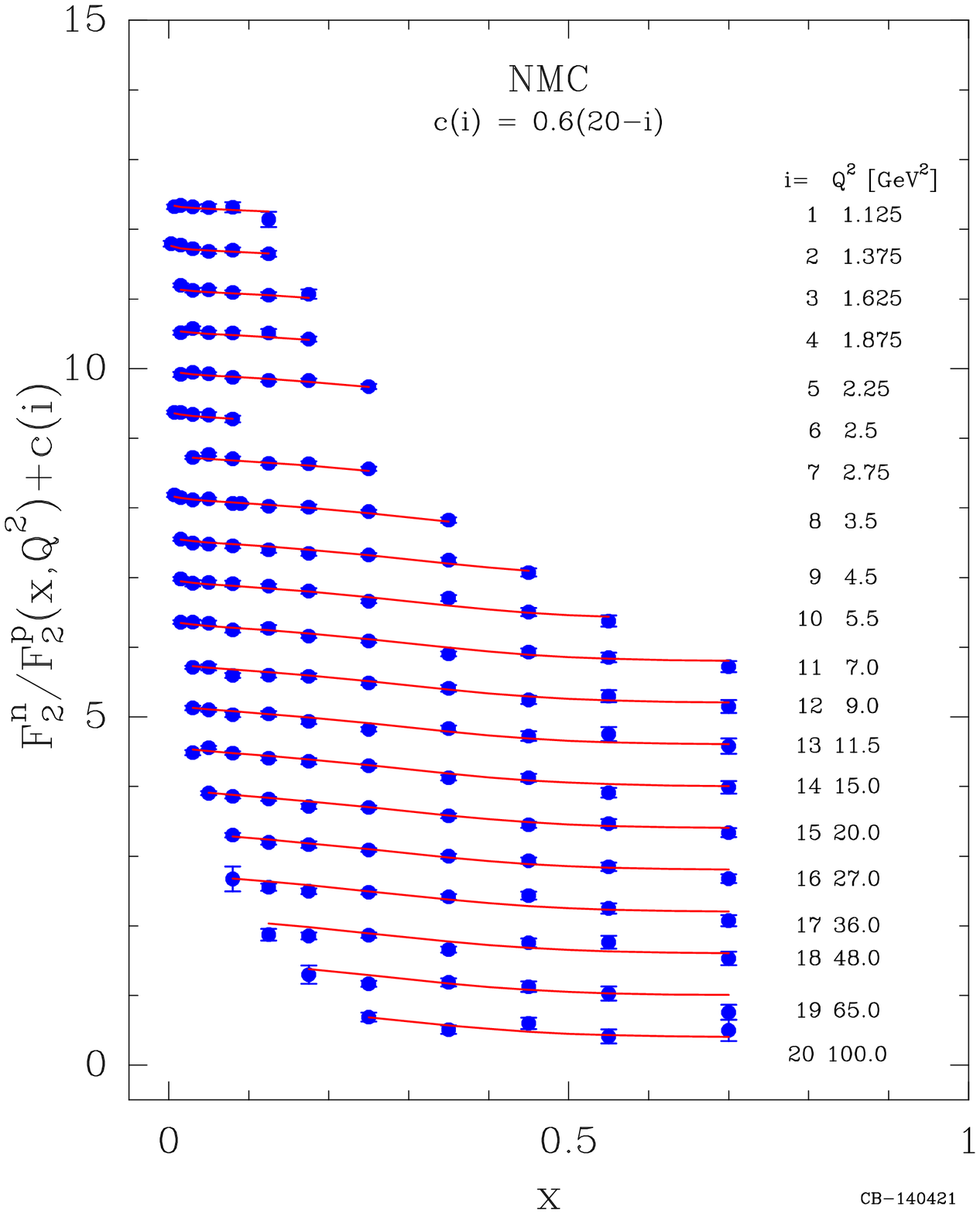}
\includegraphics[width=6.2cm]{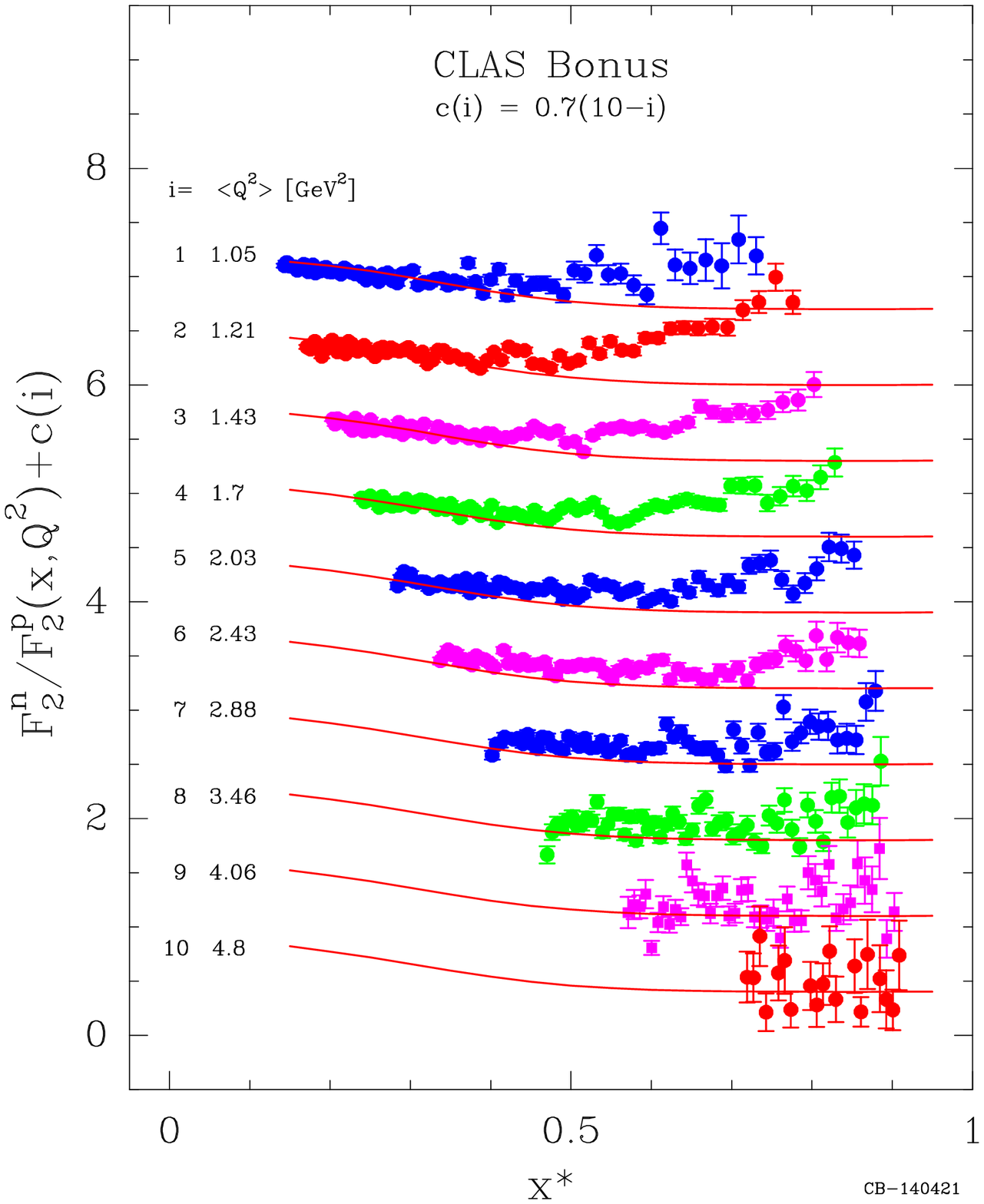}
\caption{{\it Left}~: Comparison of the data on $F_2 ^n/F_2 ^p (x,Q^2)$  from
NMC [10], with the predictions of the statistical model (solid curves).
{\it Right}~: Comparison of the data on $F_2 ^n/F_2 ^p (x,Q^2)$  from CLAS
[11], with the predictions of the statistical model (solid curves).}
\label{fig2}
\end{figure}
Another rather interesting physical quantity is the neutron $F_2 ^n$ structure
function and in particular the ratio $F_2 ^n/F_2 ^p (x,Q^2)$ which provides
strong
contraints on the PDF of the nucleon. For example the behavior of this ratio at
large $x$ is directly related to the ratio of the $d$ to $u$ quarks in the
limit $x \to 1$, a long standing-problem for the proton structure. We show in
Fig.~\ref{fig2} the results of two experiments, NMC ({\it Left}) which is very
accurate and covers a reasonnable $Q^2$ range up to $x = 0.7$ and CLAS ({\it
Right}) which covers a smaller $Q^2$ range up to larger $x$ values, both are
fairly well described by the statistical approach. Several comments are in
order. In the small $x$ region this ratio, for both cases, tends to 1 because
the structure functions are dominated by sea quarks driven by our universal
diffractive term. In the high $x$ region dominated by valence quarks, the NMC
data suggest that this ratio goes to a value of the order of 0.4 for $x$ near
1, which corresponds to the value 0.16 for $d(x)/u(x)$ when $x \to 1$, as found
in the statistical approach \cite{bbs4}. The CLAS data at large $x$ cover the
resonance region of the cross section and an important question is whether
Bloom-Gilman duality holds as well for the neutron as it does for the proton.
We notice that the predictions of the statistical approach suggest an
approximate validity of this duality, except for some low $Q^2$ values.
A better precision and the extension of this experiment with the 12GeV
Jefferson Lab will certainly provide even stronger constraints on PDFs up to $x
\simeq 0.8$.

\begin{figure}[ht]
\centering
\includegraphics[width=5.0cm]{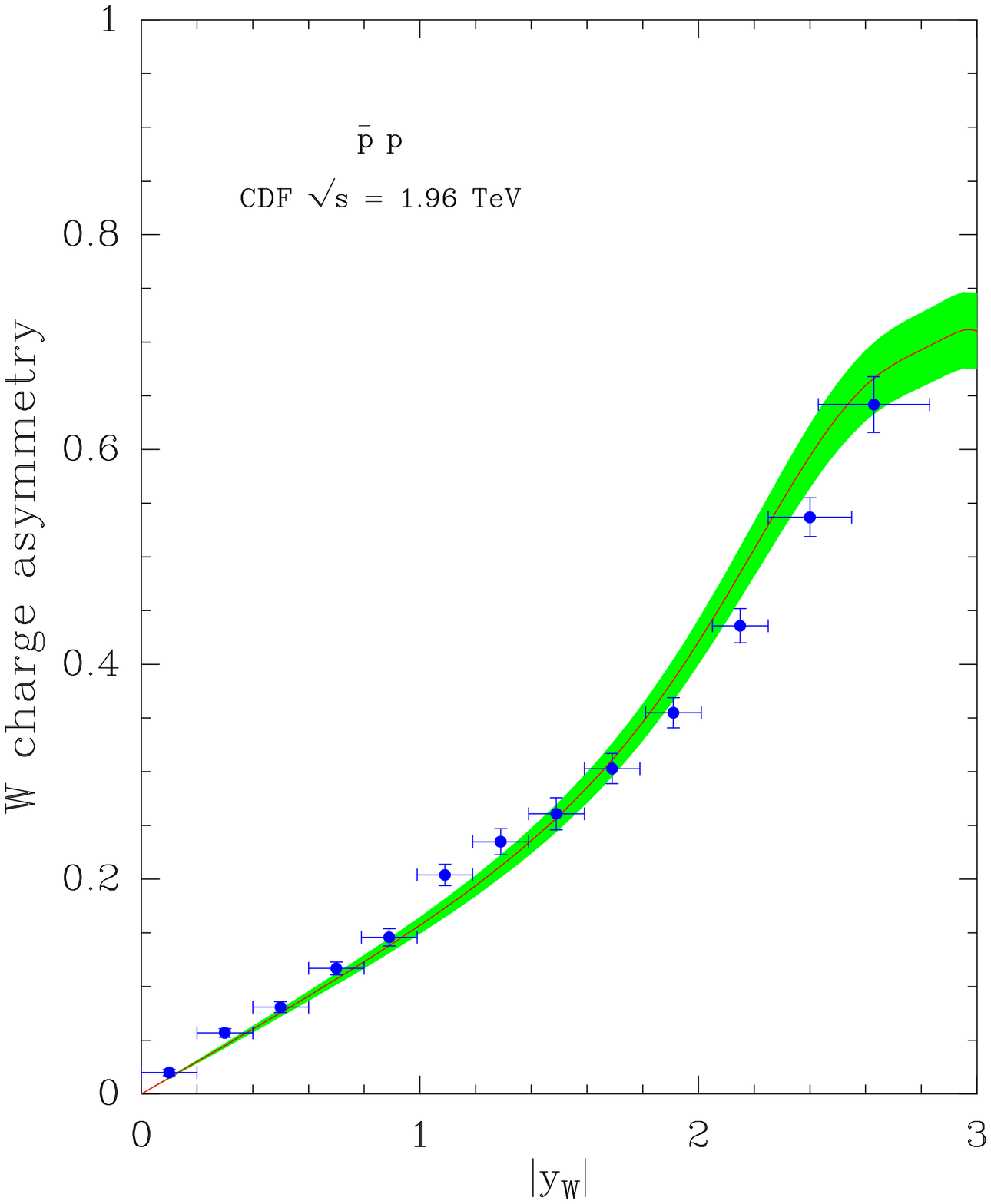}
\includegraphics[width=7.3cm]{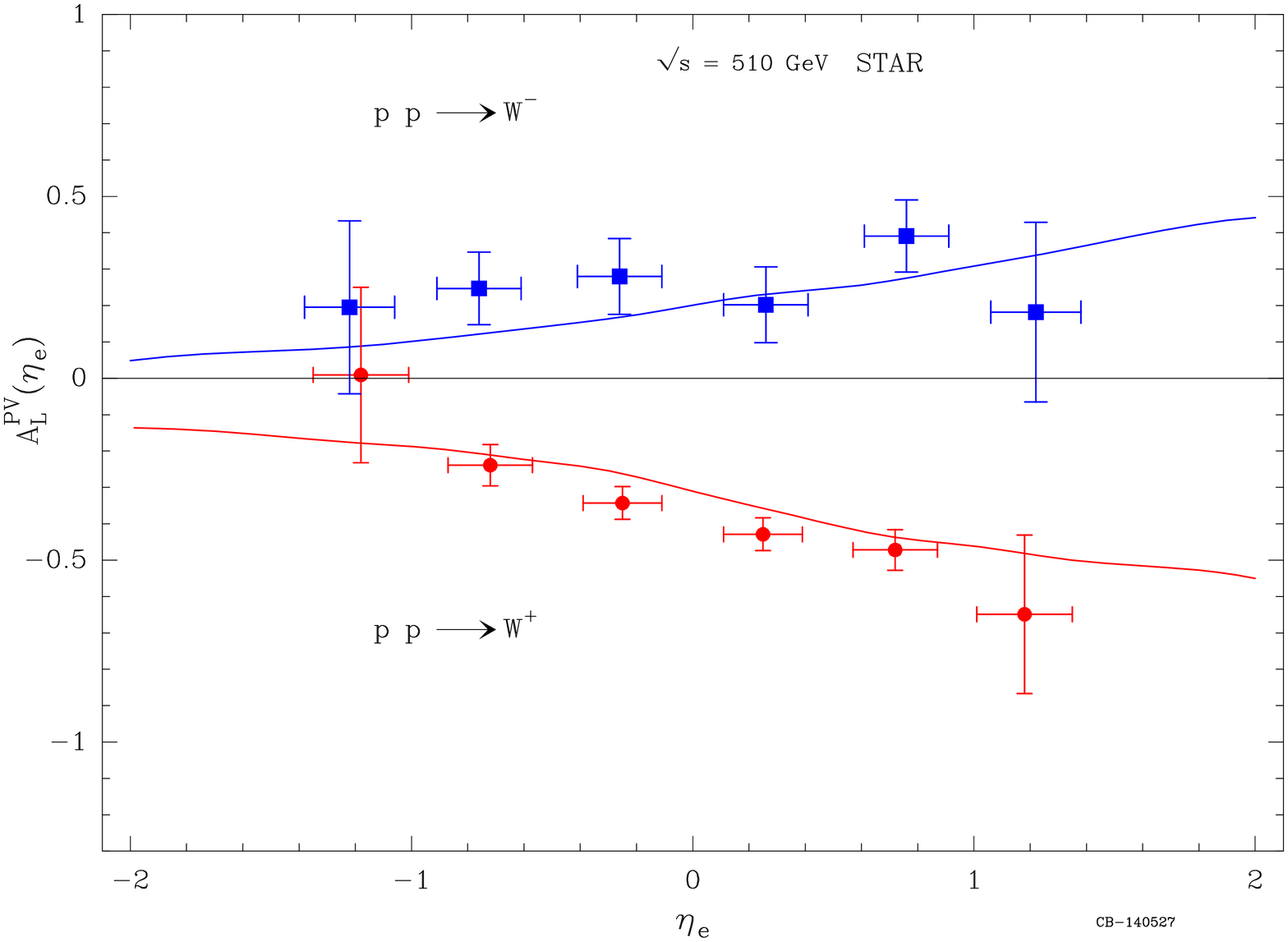}
\caption{{\it Left}~: The measured W production charge asymmetry from CDF
\cite{cdf}
versus the W rapidity $y_W$ and the prediction from the statistical
approach (solid line). The shaded band represents the uncertainties with a CL
of 68\%.
{\it Right}~: The measured parity-violating helicity asymmetries $A_L^{PV}$
for charged-lepton
production at RHIC-BNL from STAR \cite{star}, through production and decay of
$W^{\pm}$
versus $y_e$, the charged-lepton rapidity. The solid curves are the predictions
from the
statistical approach.}
\label{fig3}
\end{figure}
Let us now turn to the very interesting process of $W^{\pm}$ production in
hadronic collisions. We recall that the differential cross section in $pp$
collision $\sigma^{W^{\pm}} (y)$, where $y$ is the rapidity of the $W^{\pm}$,
can be computed directly from the Drell-Yan process dominated by
quark-antiquark fusion, $u \bar d \to W^+$ and $\bar u d \to W^-$.

\noindent The charge asymmetry defined as
\begin{equation}
A(y) = [\sigma^{W^+} (y) - \sigma^{W^-} (y)]/[\sigma^{W^+} (y) + \sigma^{W^-}
(y)]~,
\label{eq4}
\end{equation}
contains valuable information on the light quarks distributions inside the
proton and in particular on the ratio down-to-up quark. A direct measurement of
this asymmetry has been achieved by CDF at FNAL-Tevatron \cite{cdf} and the
results are shown in Fig.~\ref{fig3} ({\it Left}). The agreement with the
predictions of the statistical approach is good and we note that in the
high-$y$ region, $A(y)$ tends to flatten out, following the behavior of the
predicted $d(x)/u(x)$ in the high-$x$ region.\\
Next we consider the process $\overrightarrow p p\to W^{\pm} + X \to e^{\pm} +
X$, where the arrow denotes a longitudinally polarized proton and the outgoing
$e^{\pm}$ have been produced by the leptonic decay of the $W^{\pm}$ boson. The
helicity asymmetry is defined as
\begin{equation}
A_L^{PV} = \frac{d\sigma_+ - d\sigma_-}{d\sigma_+  + d\sigma_-}~.
\label{AL}
\end{equation}
Here $\sigma_h$ denotes the cross section where the initial proton has helicity
$h$. It was measured recently at RHIC-BNL \cite{star} and the results are shown
in Fig.~\ref{fig3} ({\it Right}). As explained in Ref. [14], the $W^-$
asymmetry is very sensitive to the sign and magnitude of $\Delta \bar u$, so
this is another successful result of the statistical approach.


\begin{figure}[ht]
\centering
\includegraphics[width=8.5cm]{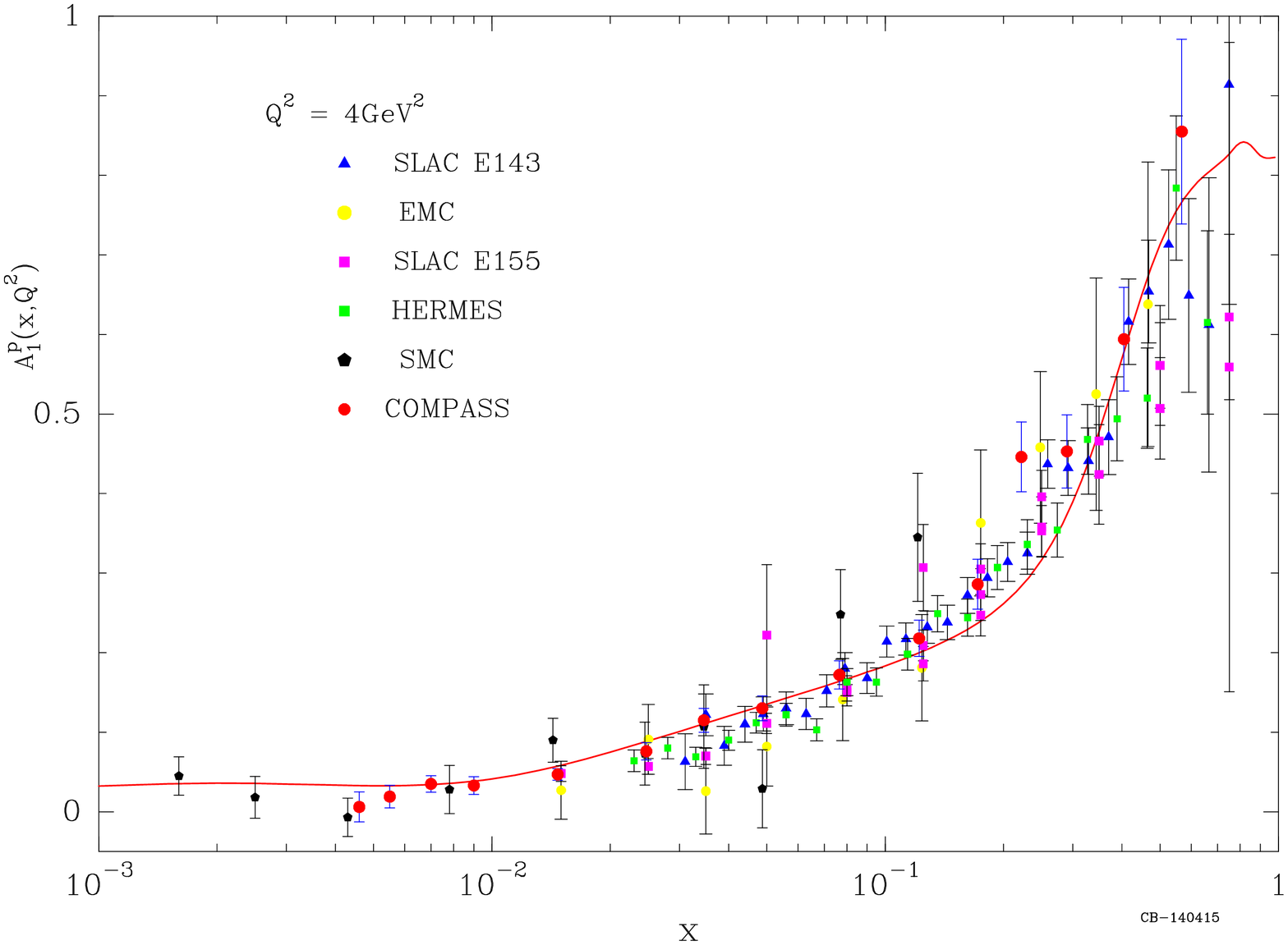}
\includegraphics[width=8.5cm]{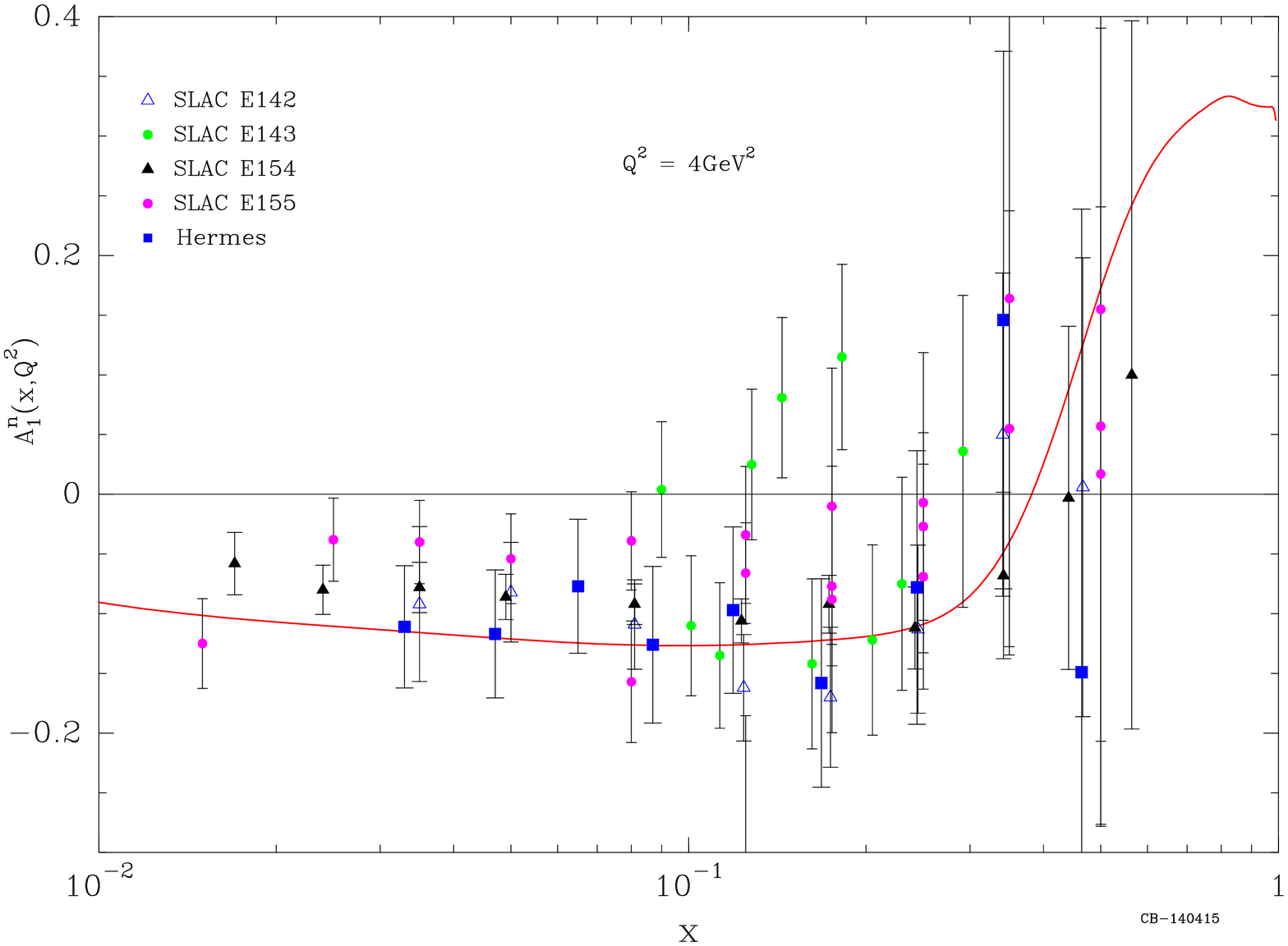}
\caption{{\it Top}~: Comparison of the world data on $A_1^p(x,Q^2)$ at
$Q^2= 4 \mbox{GeV}^2$, with the result of the statistical approach.
{\it Bottom}~: Comparison of the world data on $A_1^n(x,Q^2)$ at $Q^2= 4
\mbox{GeV}^2$,
with the result of the statistical approach.}
\label{fig4}
\end{figure}
Finally we turn to the important issue concerning
the asymmetries $A_1^{p,d,n}(x,Q^2)$, measured in polarized DIS.
We recall the definition of the asymmetry $A_1(x,Q^2)$, namely
\begin{equation}
A_1(x,Q^2)= \frac{(g_1(x,Q^2)-\gamma^2(x,Q^2)g_2(x,Q^2))2x[1+R(x,Q^2)]}
{[1+\gamma^2(x,Q^2)]F_2(x,Q^2)}~,
\label{26}
\end{equation}
where $g_{1,2}(x,Q^2)$ are the polarized structure functions,
$\gamma^2(x,Q^2)=4M^2x^2/Q^2$ and $R(x,Q^2)$ is the ratio between the
longitudinal and transverse photoabsorption cross sections. We display in
Fig.~\ref{fig4} the world data on $A_1^{p,n}(x,Q^2)$ at $Q^2= 4 \mbox{GeV}^2$,
with the results of the statistical approach.

Note that these asymmetries do NOT reach 1 when $x \to 1$ as required by the
counting rules prescription, which we don't impose.\\
Finally one important outcome of this new analysis of DIS data in the framework
of the statistical approach, is the discovery of a large gluon helicity
distribution.
When this talk was delivered, we had obtained a preliminary determination of
$x\Delta G(x,Q^2)$ which has been improved very recently and for more details
we refer the reader to Ref.~[15].

\section{Transverse momentum dependence of the parton distributions}
The parton distributions $p_i(x,k^2_T)$ of momentum $k_T$, must obey the
momentum sum rule
$\sum_i \int_0^1dx \int x p_i(x,k^2_T) dk^2_T = 1$. In addition it must also
obey
the transverse energy sum rule $\sum_i \int_0^1dx \int
p_i(x,k^2_T)\frac{k^2_T}{x}dk^2_T =M^2 $.
{}From the general method of statistical thermodynamics we are led to put
$p_i(x,k^2_T)$ in correspondance with the following expression
$\exp({\frac{-x}{\bar{x}}}+{\frac{-k^2_T}{x \mu^2}})$~,
where $\mu^2$ is a parameter interpreted as the transverse temperature.
 So we have now the main elements for the extension to the TMD of the
statistical PDF. We obtain in a natural way the Gaussian shape with {\bf no}
$x,k_T$ factorization,
because the quantum statistical distributions for quarks and antiquarks read in
this case
\begin{equation}
xq^{h}(x,k_T^{2})=\frac{F(x)}{\exp(x-X^{h}_{0q})/\bar{x}+1}
\frac{1}{\exp(k^2_T/x\mu^2-Y^{h}_{0q})+1}~,
\end{equation}
\begin{equation}
x\bar{q}^{h}(x,k_T^{2})=\frac{{\bar
 F}(x)}{\exp(x+X^{-h}_{0q})/{\bar{x}}+1}
\frac{1}{\exp(k^2_T/x\mu^2+Y^{-h}_{0q})+1}~.
\end{equation}
Here
$F(x) = \frac{A x^{b-1}X^{h}_{0q}}{\ln{(1 + \exp{Y^{h}_{0q}})}\mu^2}
=\frac{A x^{b-1}}{k\mu^2}$,
where $Y^h_{0q}$ are the thermodynamical potentials chosen such that
$\ln{(1 + \exp{Y^{h}_{0q}})}=k X^h_{0q}$,
in order to recover the factors $X^h_{0q}$ and $(X^h_{0q})^{-1}$, introduced
earlier.\\
Similarly for $\bar q$ we have $\bar F(x)= \bar A x^{2b-1}/k\mu^2$. The
determination of the 4 potentials $Y^h_{0q}$ can be achieved with the choice
$k=3.05$.
Finally $\mu^2$ will be obtained from the transverse energy sum rule and one
finds $\mu^2=0.198\mbox{GeV}^2$. Detailed results are shown in Refs.~[2,3].
Before closing we would like to mention an important point.
So far in all our quark or antiquark TMD distributions, the label "`$h$"'
stands for the
helicity along the longitudinal momentum and not along the direction of the
momentum, as normally
defined for a genuine helicity. The basic effect of a transverse momentum $k_T
\neq 0$ is the Melosh-Wigner rotation, which mixes the
components $q^{\pm}$ in the following way
$q^{+MW}= \cos^2\theta ~q^+ + \sin^2\theta ~q^- ~~~\mbox{and}~~~q^{-MW}=
\cos^2\theta ~q^- + \sin^2\theta ~q^+$, where for massless partons,
$\theta = \arctan{(\frac{k_T}{p_0 +p_z})}$, with $p_0 = \sqrt{k_T^2 +p_z^2}$.
It vanishes when either $k_T =0$ or $p_z$, the quark longitudinal momentum,
goes to infinity.
Consequently $q = q^+ + q^-$ remains unchanged since $q^{MW}=q$,
 whereas we have $\Delta q^{MW}= (\mbox{cos}^2\theta - \mbox{sin}^2\theta)
\Delta q$.
\section*{Acknowledgments}

JS is grateful to the organizers of this very successful QCD Evolution
Workshop, for their warm hospitality at Santa Fe and for providing a generous
financial support.


\end{document}